\documentstyle[12pt]{article}
\begin{document}
\title{An effective relaxation-time approach to collisionless quark-gluon plasma
\thanks{Supported by the National Natural Science Foundation of China Grant
No. 19805003
Email address  zhengxp@iopp.ccnu.edu.cn}}
\author{{Zheng Xiaoping$^{1 \hskip 2mm 2}$\ \ Li Jiarong$^{2}$\ \ Liu
Lianggang$^1$\ \ Guo Shuohong$^{1}$}\\
{\small 1 Department of Physics,Zhongshan University,Guangzhou510275,P.R.China}\\
{\small 2 The Institute of Particle Physics, Huazhong Normal
University,Wuhan430079,P.R.China}}
\maketitle 
%%%%%%%%%%%%%%%%%%%%%%%%%%%%%%%%%%%%%%%%%%%%%%%%%%%%%%%%%%%%%%%%%%%%
\vskip-9.5cm
\hskip11cm{\large HZPP-9905}

\hskip11cm{\large July 25, 1999}

\vskip8.5cm

\begin{abstract}
\begin{minipage}{120mm}
We present an effective relaxation-time theory to study the collisionless
quark-gluon plasma. Applying this method we calculate the damping rate to 
be of order $g^2T$
and find plasmon scattering is the damping mechanism. The damping
for the transverse mode is stronger than the longitudinal one.
\end{minipage}
\end{abstract}
\vskip 0.9cm

The collisional term in a transport equation is important for a many particles
system. We see the transport properties are related to collisions, for example,
the damping in plasma. Up to now, the collisional terms have not been resolved
in quark-gluon plasma(QGP). We often study the case of collisionless plasma. Although
it is well-known that the collective excitations make no contribution to
transport processes for the ultra-relativistic plasma, we should believe the
interactions of collective modes play vital role in the collisionless plasma.
Here, we just study the damping due to the interactions.

We start from the following kinetic equations for QGP\cite{r1}
\begin{equation}
p^\mu D_\mu Q_{\pm}({\bf p},x)\mp {g\over
2}p^\mu\partial^\nu_p\{F_{\mu\nu}(x),Q_\pm ({\bf p},x)\}=0,
\end{equation}
\begin{equation}
p^\mu {\cal D}_\mu G({\bf p},x)-{g\over 2}p^\mu\partial^\nu_p\{{\cal
F}_{\mu\nu}(x),G({\bf
p},x)\}=0,
\end{equation}
where the calligraphic letters represent the corresponding operators in
adjoint representation of SU(3). Now, we separate the density fluctuations
from the mean parts:$Q=\langle Q\rangle+Q^T, G=\langle G\rangle+G^T$, $\langle\
\rangle$ denotes the average over random phases\cite{r2}. 
 Assuming the fluctuations to be weak,
$p\sim gT, A_\mu\sim T, i\partial_\mu\sim gT$,
we neglect the Abelian-like coupling of the fluctuations because the coupling
is of higher
order $g$ than the non-Abelian one.
Thus we write the equations
for the fluctuations
\begin{eqnarray}
p^\mu\partial_\mu Q^T_\pm\mp gp^\mu F_{\mu\nu}\partial_p^\nu Q^T_+=
-igp^\mu\left ([A_\mu , Q^T_+]-\langle [A_\mu , Q^T_+]\rangle\right ),
\end{eqnarray}
%\begin{eqnarray*}
%p^\mu\partial_\mu Q^T_-+gp^\mu F_{\mu\nu}\partial_p^\nu Q^T_-=
%-igp^\mu\left ([A_\mu , Q^T_-]-\langle [A_\mu , Q^T_-]\rangle\right ),
%\end{eqnarray*}
\begin{eqnarray}
p^\mu\partial_\mu G^T-gp^\mu{\cal F}_{\mu\nu}\partial_p^\nu G^T=
-igp^\mu\left ([{\cal A}_\mu , G^T]-\langle [{\cal A}_\mu , G^T]\rangle\right ),
\end{eqnarray}

Effectively, we regard the nonlinear parts in the equations as 'collisional
terms', which they represent the interactions of collective modes.
We now define the effective collisional frequencies $\nu_q,\nu_{\bar q}$
and $\nu_g$ for quarks, antiquarks and gluons. In the relaxation-time
approach, we have identities
\begin{eqnarray}
p^\mu u_\mu\nu_qQ^T_\pm (k)=\int{{\rm d}k_1{\rm d}k_2\over (2\pi)^8}
{\rm Im}\left (gp^\mu([A_\mu(k_2),Q^T_\pm (k_1)]-\langle [A_\mu(k_2),Q^T_\pm(k_1)]\rangle)\right ),
\end{eqnarray}
%\begin{eqnarray*}
%p^\mu u_\mu\nu_qQ_-(k)=\int{{\rm d}k_1{\rm d}k_2\over (2\pi)^8}
%{\rm Im}\left (gp^\mu([A_\mu(k_2),Q^T_-(k_1)]-\langle [A_\mu(k_2),Q^T_-(k_1)]\rangle)\right ),
%\end{eqnarray*}
\begin{eqnarray}
p^\mu u_\mu\nu_qG^T(k)=\int{{\rm d}k_1{\rm d}k_2\over (2\pi)^8}
{\rm Im}\left (gp^\mu([{\cal A}_\mu(k_2),G^T(k_1)]-\langle [{\cal A}_\mu(k_2),G^T(k_1)]\rangle)\right
).
\end{eqnarray}
We can derive all orders of density fluctuations from the asymptotic method
developed from weak-turbulent theory\cite{r3}. Here we only keep the first two orders.
In the plasma rest frame,  taking temporal gauge condition, we easily obtain
{\small
\begin{eqnarray}
\nu_q\int {{\rm d}{\bf v}\over 4\pi}\langle Q^{(1)}_+(k){\bf v\cdot A}(k')\rangle
=g\int{{\rm d}k_1{\rm d}k_2\over (2\pi)^8}\delta(k-k_1-k_2)
{\rm Im}\int{{\rm d}{\bf v}\over 4\pi}\langle [{\bf v\cdot A}(k_2),Q^{(2)}_+(k_1)]{\bf v\cdot A}\rangle,
\end{eqnarray}
\begin{eqnarray}
\nu_{\bar q}\int {{\rm d}{\bf v}\over 4\pi}\langle Q^{(1)}_-(k){\bf v\cdot A}(k')\rangle
=g\int{{\rm d}k_1{\rm d}k_2\over (2\pi)^8}\delta(k-k_1-k_2)
{\rm Im}\int{{\rm d}{\bf v}\over 4\pi}\langle [{\bf v\cdot A}(k_2),Q^{(2)}_-(k_1)]{\bf v\cdot A}\rangle,
\end{eqnarray}
\begin{eqnarray}
\nu_g\int {{\rm d}{\bf v}\over 4\pi}\langle G^{(1)}(k){\bf v\cdot{\cal A}}(k')\rangle
=g\int{{\rm d}k_1{\rm d}k_2\over (2\pi)^8}\delta(k-k_1-k_2)
{\rm Im}\int{{\rm d}{\bf v}\over 4\pi}\langle [{\bf v\cdot {\cal A}}(k_2),G^{(2)}(k_1)]{\bf v\cdot {\cal A}}\rangle.
\end{eqnarray}
}
One should note that for a baryonless plasma the numbers of quarks and antiquarks
are equal to one another and $\nu_q=\nu_{\bar q}$
(We will omit variable $k_i$ in any a quanta when the equations are mistaken in next text).
%Firstly, we write the equation as the form of mean-field correlations in
%order to get $\nu_q$,
%\begin{eqnarray*}
%&\ &\nu_q\int{{\rm d}{\bf v}\over 4\pi}{\omega\over\omega-{\bf k\cdot v}}\langle {\bf v\cdot Av\cdot A}\rangle\\
%&=&g^2{\rm Im}\int{{\rm d}k_1{\rm d}k_2{\rm d}k_3\over (2\pi)^{12}}\delta (k-k_1-k_2-k_3)
%\int{{\rm d}{\bf v}\over 4\pi}{1\over \omega_3+\omega_2-{\bf k_3\cdot v}-{\bf k_2\cdot v}}{\omega_3\over\omega_3-{\bf k_3\cdot v}}\\
%&\ &\left (\langle [{\bf v\cdot A},[{\bf v\cdot A},{\bf v\cdot A}]]{\bf v\cdot A}\rangle
%- \langle [{\bf v\cdot A},\langle [{\bf v\cdot A},{\bf v\cdot A}]\rangle ]{\bf v\cdot A}\rangle\right ).
%\end{eqnarray*}
Therefore $\nu_q$ is expressed as
{\tiny
\begin{eqnarray}
\nu_q&=&
g^2 \int{{\rm d}k'{\rm d}k_1{\rm d}k_2{\rm d}k_3\over (2\pi)^{12}}\delta (k-k_1-k_2-k_3)\\
&\times&{\int{{\rm d}{\bf v}\over 4\pi}{\rm Im}({1\over \omega_3+\omega_1-{\bf k_3\cdot v}-{\bf k_1\cdot v}})
({\omega_3\over\omega_3-{\bf k_3\cdot v}}-{\omega_1\over\omega_1-{\bf k_1\cdot v}})
f_{aed}f_{bce}\langle {\bf v\cdot A}_a{\bf v\cdot A}_b\rangle\langle{\bf v\cdot A}_c{\bf v\cdot A}_d\rangle
\over
\int{{\rm d}k'{\rm d}{\bf v}\over 4\pi}{\omega\over\omega-{\bf k\cdot v}}\langle 2{\rm tr}({\bf v\cdot Av\cdot A})\rangle }
\end{eqnarray}
}$\nu_g$ is obtained likewise
{\tiny
\begin{eqnarray}
\nu_g&= &
Ng^2\int{{\rm d}k'{\rm d}k_1{\rm d}k_2{\rm d}k_3\over (2\pi)^{12}}\delta (k-k_1-k_2-k_3)\\
&\times&{\int{{\rm d}{\bf v}\over 4\pi}{\rm Im}({1\over \omega_3+\omega_1-{\bf k_3\cdot v}-{\bf k_1\cdot v}})
({\omega_3\over\omega_3-{\bf k_3\cdot v}}-{\omega_1\over\omega_1-{\bf k_1\cdot v}})
f_{aed}f_{bce}\langle {\bf v\cdot A}_a{\bf v\cdot A}_b\rangle\langle{\bf v\cdot A}_c{\bf v\cdot A}_d\rangle
\over
\int{{\rm d}k'{\rm d}{\bf v}\over 4\pi}{\omega\over\omega-{\bf k\cdot v}}\langle {\rm tr}({\bf v\cdot {\cal A}v\cdot {\cal A}})\rangle }
\end{eqnarray}
}
We can determine the damping rate $\gamma$ from the
following derivation,
\begin{eqnarray*}
&\ &N_f\nu_q\langle (Q^{(1)}_+-Q^{(1)}){\bf v\cdot A}\rangle+2\nu_g\tau_a\langle{\rm tr}(T_aG^{(1)}){\bf v\cdot A}\rangle\\
&=&\gamma\left (N_f\langle (Q^{(1)}_+-Q^{(1)}){\bf v\cdot A}\rangle+2\tau_a\langle{\rm tr}(T_aG^{(1)}){\bf v\cdot A}\rangle\right ),
\end{eqnarray*}
and then
\begin{eqnarray}
\gamma={N_f\nu_q\int{\rm d}^3p\partial^0_pQ^R+N\nu_g\int{\rm d}^3p\partial^0_pG^R
\over N_f\int{\rm d}^3p\partial^0_pQ^R+N\int{\rm d}^3p\partial^0_pG^R}.
\end{eqnarray}

%Obviously, we can prove $\gamma=\nu_q=\nu_g$ from the above formulae.
Of course,  for the case of  white noise with same color inherent we have
{\tiny
\begin{eqnarray}
\gamma=g^2N{
\int{{\rm d}k_2\over (2\pi)^4}{{\rm d}{\bf v}\over 4\pi}\pi\delta[\omega_2-\omega-({\bf k}_2-{\bf k})\cdot {\bf v}]
({\omega_2\over\omega_2-{\bf k}_2\cdot{\bf v}}-{\omega\over\omega-{\bf k\cdot v}})
({({\bf k}_2\cdot{\bf v})^2\over {\bf k}^2_2}\langle A_l^2(k_2)\rangle
 +{({\bf k}_2\times{\bf v})^2\over {\bf k}^2_2}\langle A_t^2(k_2)\rangle)
 \over
 \int {{\bf v}\over 4\pi}{\omega\over\omega-{\bf k\cdot v}}
 }
 \end{eqnarray}
} where $l$ and $t$ respectively denote the longitudinal and transverse
 components of the field.
 Now we discuss the specific results with respect to $\gamma$.
We know
\begin{eqnarray*}
{\bf A\cdot A}(\omega, {\bf k})=A^2_l(\omega, {\bf k})+A^2_t(\omega, {\bf k}),
\end{eqnarray*}
\begin{eqnarray*}
\langle A^2(\omega,{\bf k})\rangle=I({\bf k}){\pi\over\omega^2}[\delta(\omega-\omega({\bf k}))+\delta(\omega+\omega({\bf k}))],
\end{eqnarray*}
 where $I({\bf k})$ represents the correlation intensity  of collective fields
  with frequencies $\omega({\bf k})$ and $-\omega({\bf k})$, here we think
only the positive frequency mode exists and $I=4\pi T$ when
  the plasma is equilibrium. We discuss longwavelenth modes for three cases\\
  (i) The case of longitudinal wave: $A^2_l=A^2, A^2_t=0$, 
  the dispersive relation is $\omega_l({\bf k})=\omega^2_p+{3\over 5}{\bf
k}^2$,  then we obtain
  \begin{eqnarray*}
  \gamma_l=0.01g^2T
  \end{eqnarray*}
  (ii)The case of transverse wave: $A^2_l=0, A^2_t=A^2$, the dispersive
relation is
  $\omega_t({\bf k})=\omega^2_p+{6\over 5}{\bf k}^2$,  we have
  \begin{eqnarray*}
  \gamma_t= 0.43g^2T
  \end{eqnarray*}
  (iii)We can also calculate $\gamma$ for the mixed case with $A^2_l=A^2_t$.
  \begin{eqnarray*}
  \gamma_{l+t}= 0.21g^2T
  \end{eqnarray*}

In fact, we can calculate $\gamma$ for more general case if the oscillational
direction of the collective modes with respect to wave vector is given.
Although we have not treated general case, we see that the damping
for a transverse mode is much stronger
than the longitudinal one and the damping rate is on the increase as
polarized field direction deviation from the direction of wave motion.

One has been applying 'hard thermal loop" method\cite{r4} to discuss the
damping rate in high temperature plasma. The order do coincides with the one
for equilibrium which we study above in the framework of kinetic theory.
It seem to imply a obvious connection to classical nonlinear effect
from 'hard thermal loop'. Moreover, our study can be extended to out of
equilibrium.

%The above results are obtained under the condition of equilibrium fluctuations.
%We see that the thermal fluctuating intensity is small. However, one predicts
%to produce turbulent plasma in heavy ion collisions\cite{}, then we expect
%that the collisionless effect become more important since the fluctuations
%will grow enormously.

The damping do not arise from collisional process though we apply
the relaxation-time method. Here we easily analyzes the physical
mechanism of the damping from the
above calculations. The previous equations contain the field powers
from $({\bf v\cdot A})^2$ to$({\bf v\cdot A})^4$.
It means the existing of Cherenkov process, decay process and
scattering process in  interactions of plasmons(collective modes)
 with the particles.
 It is possible that
all these processes cause particles and plasmon(collective) to exchange
the energy and momentum each other. However, we know the Cherenkov process
cannot be produced due to $\omega>|{\bf k}|$ in linear response
approximation\cite{r5}. In addition, the decay of plasmon has been proved vanish
since $\int{{\rm d}{\bf v}\over 4\pi}{\bf vvv}f(v)=0$. So we emphasize
the plasmon scattering is the mechanism of collisionless damping, which
represents nonlinear interaction.

\end{document}